\documentclass[]{article}

 \usepackage{amsmath,amsthm}
 \usepackage{amssymb}
\usepackage{stmaryrd}

\usepackage{tikz}
\usetikzlibrary{arrows,automata}
\tikzstyle{block}=[draw opacity=0.7,line width=1.4cm]

\newcommand{\noop}[1]{}

\newcommand{\query}{\ensuremath{Q}}
\newcommand{\eval}[2]{\ensuremath{\llbracket #1 \rrbracket_{#2}}}

\newcommand{\core}[1]{\ensuremath{#1^c}}

\newcommand{\langCQ}{\ensuremath{CQ}}

\newcommand{\langTwoCQ}{\ensuremath{2CQ_{\sigma,\alpha}}}
\newcommand{\constants}{\ensuremath{\mathcal{C}}}
\newcommand{\containedIn}{\mathrel{\sqsubseteq}}
\newcommand{\maxContainedIn}{\mathrel{\sqsubseteq_m}}
\newcommand{\propContainedIn}{\mathrel{\sqsubset}}
\newcommand{\distCont}{\ensuremath{\delta_\sqsubseteq}}

\newcommand{\graph}{\ensuremath{G_{\sigma,\alpha}}}
\usepackage{fullpage}
\usepackage{times}

\newtheorem{theorem}{Theorem}[section]
\newtheorem{lemma}[theorem]{Lemma}
\newtheorem{proposition}[theorem]{Proposition}

\newtheorem{definition}[theorem]{Definition}

\newtheorem{example}[theorem]{Example}

\begin{document}

\title{On the feasibility of semantic query metrics}

\author{George Fletcher$^1$, Peter Wood$^2$, Nikolay Yakovets$^1$}

\date{$^1$Eindhoven University of Technology, The Netherlands \\ 
$^2$Birkbeck, University of London \\
{\small \texttt{$\{$g.h.l.fletcher, n.yakovets$\}$@tue.nl}, 
\texttt{p.wood@bbk.ac.uk}}}

\maketitle
%

\begin{abstract}
We consider the problem of defining semantic metrics for relational database queries.   Informally, a semantic query metric for a query language $L$ is a metric function $\delta:L\times L\to \mathbb{N}$ where $\delta(Q_1, Q_2)$ represents the length of a shortest path between queries $Q_1$ and $Q_2$ in a graph. In this graph, nodes are queries from $L$, and edges connect semantically distinct queries where one query is maximally semantically contained in the other.
Since query containment is undecidable for first-order queries, we focus on the simpler language of conjunctive queries. We establish that defining a semantic query metric is impossible even for conjunctive queries. Given this impossibility result, we identify a significant subclass of conjunctive queries where such a metric is feasible, and we establish the computational complexity of calculating distances within this language.
\end{abstract}




\section{Introduction}
\label{sec:intro}

Defining distances between queries plays an important role in the study of data management systems.
Example applications giving rise to notions of query distance include
workload compression and clustering \cite{jain2018b,kul2018,xie2018};
    view selection, index selection and, in general, physical design  \cite{Afrati,ChaudhuriGN02,ChirkovaY12,GebalyA08,graefe2012,MozafariGY15};
    statistical synopses construction (e.g., histogram tuning) \cite{ChaudhuriGN02},
    approximate aggregate query answering \cite{ChaudhuriGN02};
    workload forecasting and monitoring \cite{jain2018a,MaAHMPG18,MOREAU2022};
    query recommendation \cite{jain2018a,jain2018b};
    benchmark design \cite{jain2018a,jain2018b};
    query debugging \cite{HuGSRY24,jain2018a,jain2018b};
    workload error classification \cite{jain2018a,jain2018b}; 
    database forensics \cite{jain2018a,jain2018b,MOREAU2022}; and
    data systems education \cite{HuGSRY24,koberlein2024,RoyGHMMS024,YangGHMMRS24}.

The majority of studies
have defined distance in terms of the syntax of
queries, either using the query-strings themselves or some derivative thereof,
such as physical query execution plans.  
Syntactic and instance distance measures have also been studied 
in the database and logics communities, e.g., 
\cite{ArieliDB07,Libkin04,RamonB98,RamonB01}.
Alternatively, some studies have taken into
consideration operational behaviour such as query frequency and workload co-occurence in the definition of distance, e.g.,  \cite{MaAHMPG18}. Furthermore, heuristic machine learning approaches have been studied, e.g., \cite{tang}.

Distance functions studied in the literature often contain some indirect proxy of the semantics of queries,
e.g., queries with shared physical access paths will potentially have some
semantic similarity.  To our knowledge, the semantic behaviour of queries have
not been directly taken into consideration in the study of query distance
(where ``behaviour'' means input/output behaviour).  
This is surprising, as many workload analytics tasks rely on grouping queries
with similar behaviors.

\paragraph{Contributions.}  
Given these practical observations,
in this work we consider the problem of defining semantic metrics for relational database queries.   Informally, a semantic query metric with respect to query language $L$ is a metric function $\delta:L\times L\to \mathbb{N}$ such that  $\delta(Q_1, Q_2)$ is the number of queries on a shortest path between $Q_1$ and $Q_2$ in the graph having node set $L$ and edge set $\{ \{Q, Q'\} \mid $ $Q$ and $Q'$ are semantically distinct and $Q$ is maximally semantically contained in $Q'\}$.
As containment is undecidable for first order queries, we consider the language of conjunctive queries for which containment is decidable \cite{ChandraM77}.  We establish that it is also not possible to define a semantic query metric for this language.  Given this impossibility result, we identify a non-trivial class of conjunctive queries for which this is possible and establish the complexity of computing distances within this language.  
\section{Preliminary definitions}

We fix our attention on the language of conjunctive queries. 

\paragraph*{Data}
A {\em relational schema} is a finite set of {\em relation} names $\sigma = \{R_1,
\ldots, R_k\}$, where each $R\in\sigma$ is of some fixed $arity(R)\in \{0, 1, \ldots \}$.  

Let \constants{} be a domain of constants.
A {\em $\sigma$-fact} is a tuple $R(c_1, \ldots, c_a)$ where $R\in \sigma$, $a= arity(R)$, and
$c_i\in \constants$ for each $1\leq i \leq a$.
An {\em instance} of $\sigma$ is a finite set $I$ of $\sigma$-facts.

\paragraph*{Queries}
Let $\sigma$ be a relational schema, $\mathcal{V}$ be a set of variables, and $n>0$. 
A {\em conjunctive query (\langCQ)} over $\sigma$ is a rule of the form
\begin{eqnarray*}
    (\overline{z}) &\gets & S_1(\overline{x}_1), \ldots, S_n(\overline{x}_n)
\end{eqnarray*}
where 
\begin{itemize}
    \item $S_j \in \sigma$ and $\overline{x}_j \subseteq \mathcal{V}$ is a list of variables of length $arity(S_j)$, 
        for $1\leq j \leq n$; and,
    \item $\overline{z} \subseteq \bigcup \overline{x}_j$ is a list of variables.
\end{itemize}
We call $(\overline{z})$ the {\em head} and $S_1(\overline{x}_1), \ldots, S_n(\overline{x}_n)$
the {\em body} of the query; 
we call the variables that appear in the head {\em distinguished\/} variables,
and those that appear only in the body {\em non-distinguished\/} variables;
and, we call the elements $S_j(\overline{x}_j)$ the {\em atoms} of the body.
Further, we say the query is of {\em arity} $|\overline{z}| \geq 0$.

Given query $Q = (z_1, \ldots, z_m) \gets S_1(x^1_1, \ldots, x^1_{i_1}), \ldots, S_n(x^n_1, \ldots, x^n_{i_n})$ 
and instance $I$ of $\sigma$, a function $h: \mathcal{V} \to \mathcal{C}$, is an {\em embedding of 
$Q$ in $I$} if, for each atom $S_j(x^j_1, \ldots, x^j_{i_j})$ in the body of $Q$, it holds that 
$S_j(h(x^j_1), \ldots, h(x^j_{i_j})) \in I$.  The semantics of $Q$ on $I$ is the set
\begin{eqnarray*}
    \eval{Q}{I} &=& \{(h(z_1), \ldots, h(z_m)) \mid h \textnormal{ is an embedding of } Q \textnormal{ in } I\}.
\end{eqnarray*}

\begin{example}\label{ex:instance-cq}
Assume we have the relational schema $\sigma = \{ R, L \}$, where $arity(R)=2$ and $arity(L)=1$.
Let $I$ be the following instance of $\sigma$:
\[
R(a, b), R(a, c), R(b, a), R(c, c)
\]
\[
L(a), L(b), L(c)
\]
Consider the following CQ $Q$ over $\sigma$:
\begin{eqnarray*}
(x, y) & \leftarrow & R(x, y), R(y, x), L(x), L(y) .
\end{eqnarray*}
One embedding of $Q$ in $I$ is given by the function $h_1$ which maps $x$ to $a$
and maps $y$ to $b$, while another is given by $h_2$ which maps both $x$ and $y$ to $c$.
Since there are no other embeddings, we have $\eval{Q}{I} = \{(a,b), (c,c) \}$.
\end{example}

\subsection{Maximal query containment}
The classic notions of relative semantic behaviour are query containment and equivalence.
Our semantic distance function is based on the notion of maximal query containment
which we define below.

\begin{definition} \label{def:containment}
    Let $\query_1, \query_2\in\langCQ$.
\begin{itemize}

    \item $\query_1$ is {\em contained} in $\query_2$, denoted  $\query_1 \containedIn
    \query_2$, if and only if for every instance $I$ it holds that
        $\eval{\query_1}{I} \subseteq \eval{\query_2}{I}$.
        
    \item $\query_1$ and $\query_2$ are {\em equivalent}, denoted
    $\query_1 \equiv \query_2$, if and only if $\query_1 \containedIn \query_2$
    and $\query_2 \containedIn \query_1$.

\item The {\em core} of a query $\query$ is the equivalent query
    $\core{\query}$ having the smallest body, i.e., $\core{\query}$ has the
        fewest atoms in its body of all queries equivalent to
        \query.\footnote{We say ``the'' core, as cores are computable and unique up to
        renaming of variables \cite{HellN92}.}  We also say that $\query$ is {\em minimal\/}
        if $\query = \core{\query}$ (up to variable renaming).

\end{itemize}
\end{definition}

\begin{definition} \label{def:homo}
A {\em homomorphism} from query $\query_2$
to query $\query_1$ is a function $h:\mathcal{V}\to\mathcal{V}$ such that, (1)~for each atom 
$S_j(x^j_1, \ldots, x^j_{i_j})$ in the body of $\query_2$, it is the case that 
$S_j(h(x^j_1), \ldots, h(x^j_{i_j}))$ is an atom in the body of 
$\query_1$; and, (2)~given the head $(z_1, \ldots, z_m)$ of $\query_2$, 
$(h(z_1), \ldots, h(z_m))$ is the head of $\query_1$.
We denote that such a homomorphism exists by $\query_2\to \query_1$.
\end{definition}

Note that the above definition allows a homomorphism to map
one distinguished variable to another.
As is common practice, we will also apply homomorphisms to tuples 
of variables, with the obvious meaning: i.e., if $\overline{z} = (z_1, \ldots, z_m)$,
then $h(\overline{z}) = (h(z_1), \ldots, h(z_m))$.

We have the following useful characterisation of containment.
\begin{proposition}[\cite{ChandraM77}]\label{prop:containment-homo}
$\query_1 \containedIn \query_2$ if and only if $\query_2 \to \query_1$.
\end{proposition}

\begin{example}\label{ex:homomorphism}
Let $Q_1$ be the CQ $Q$ from Example~\ref{ex:instance-cq}:
\begin{eqnarray*}
(x, y) & \leftarrow & R(x, y), R(y, x), L(x), L(y)
\end{eqnarray*}
and $Q_2$ be the CQ
\begin{eqnarray*}
(x, y) & \leftarrow & R(x, z), L(y), L(z)
\end{eqnarray*}
With the instance $I$ from Example~\ref{ex:instance-cq},
we see that $\eval{Q_2}{I} = \{ (a,a), (a,b), (a,c), (b,a), (b,b), (b,c), (c,a), (c,b), (c,c) \}$,
i.e., $\eval{Q_1}{I} \containedIn \eval{Q_2}{I}$.  In fact, $Q_1 \containedIn Q_2$ since there is
a homomorphism $h$ from $Q_2$ to $Q_1$ which maps $x$ and $y$ to themselves and maps $z$ to $y$.

If we add the atom $R(y, z)$ to $Q_1$ to give $Q_3$:
\begin{eqnarray*}
(x, y) & \leftarrow & R(x, y), R(y, x), R(y, z), L(x), L(y)
\end{eqnarray*}
we see that $Q_3$ is not minimal since it is equivalent to $Q_1$.
We could add any number of atoms of the form $R(y, w_i)$ to $Q_3$, where in each case $w_i$ 
is a new variable, showing that the same query can be represented syntactically in infinitely many ways.
\end{example}

It is well-known that:
\begin{proposition}[\cite{ChandraM77,HellN92}]\label{prop:containment-complexity}
Containment (and therefore equivalence) is NP-complete.  Furthermore, constructing the core of a query is an NP-complete problem.  
\end{proposition}

In our study, we are interested in inequivalent queries which are semantically as ``close'' as possible.  
This is captured by the notion of {\em maximal containment}.

\begin{definition} \label{def:maxContainment}
Let $\query_1, \query_2 \in \langCQ$. 
$\query_1$ is {\em maximally contained} in $\query_2$ 
(written $\query_1 \maxContainedIn \query_2$) if and only if
(1)~$\query_1 \containedIn \query_2$, (2)~$\query_2 \not\containedIn 
\query_1$, and (3) for every $\query \in \langCQ$, it is the case that
if $\query_1 \containedIn \query  \containedIn \query_2 $ then 
$\query_1 \equiv \query $ or $\query \equiv \query_2 $.
\end{definition}

In the next section, we show that, for a certain very restricted subclass of conjunctive queries,
given a query $Q$ in the class, there is no (finite) query that is maximally contained in $Q$.

\section{Non-existence results}
\label{sec:non-existence}

We now assume that our database consists of a single binary relation $E$,
which can be viewed as the edge relation of some graph.

\begin{definition}
A Boolean {\em path query} (PQ) of length $n$, denoted $P_n$, checks for 
the existence of a (directed) path of length $n$ in relation $E$, i.e.:
$$
() \gets E(z_1, z_2), E(z_2, z_3), \ldots, E(z_n, z_{n+1}) .
$$
We refer to the atom $E(z_i, z_{i+1})$ as the $i$'th atom in $P_n$.
An {\em oriented path query} (OPQ) of length $n$ generalises a path query
by allowing the $i$'th atom to be either $E(z_i, z_{i+1})$ (called a {\em forward\/} atom)
or $E(z_{i+1}, z_i)$ (called a {\em backwards\/} atom).
\end{definition}

If we denote a forwards atom by $1$ and a backwards atom by $0$,
then we can represent any OPQ of length $n$ by $O_k$,
where $k$ is a binary sequence of length $n$.  
So, e.g., the OPQ $O_{1111}$ is the same as $P_4$, the PQ of length four\footnote{Oriented 
path queries and the binary notation for them have been inspired by~\cite{HellN96}.}.

Many oriented path queries are equivalent to one another, as shown by each row in the table 
in Figure~\ref{fig:OPQ-equivalence}
(where queries are represented simply by their binary sequences).  Some of these equivalences
arise through ``reading'' the atoms in a query from right to left rather than left to right 
(the ``Reversal'' column in the table, where a blank entry means the reversal is the same 
as the original query).  Others arise as a result of some atoms being redundant
(the ``Minimal'' column in the table, where a blank entry means the original query is minimal).  
In fact, each oriented path query of length up to $4$ is equivalent to a path query of the same 
or shorter length, as shown in the table.

\begin{figure}\centering
\begin{tabular}{lll}\hline
Query & Reversal & Minimal \\
\hline
1 & 0 & \\
\hline
11 & 00 & \\
10 & 01 & 1 \\
\hline
111 & 000 & \\
110 & 100 & 11 \\
001 & 011 & 11 \\
101 & 010 & 1 \\
\hline
1111 & 0000 & \\
1110 & 1000 & 111 \\
0111 & 0001 & 111 \\
1101 & 0100 & 11 \\
1011 & 0010 & 11 \\
1001 & 0110 & 11 \\
1100 & & 11 \\
0011 & & 11 \\
1010 & & 1 \\
0101 & & 1 \\
\hline
\end{tabular}
\caption{Equivalences among oriented path queries.}
\label{fig:OPQ-equivalence}
\end{figure}

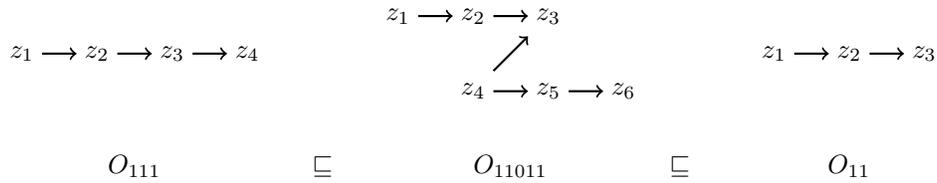
\begin{figure*}\centering
\begin{tikzpicture}[auto, swap]
\tikzstyle{vertex}=[minimum size=15pt,inner sep=0pt]
\tikzstyle{edge} = [draw,thick,->]

    \foreach \pos / \name / \label in {{(-5,9.5)/z11/$z_1$}, {(-4,9.5)/z12/$z_2$}, {(-3,9.5)/z13/$z_3$}, {(-2,9.5)/z14/$z_4$}}
        \node[vertex] (\name) at \pos {\label};
    \foreach \source / \dest in {z11/z12, z12/z13, z13/z14}
        \path[edge] (\source) -- (\dest);
        
    \foreach \pos / \name / \label in {{(0,10)/z1/$z_1$}, {(1,10)/z2/$z_2$}, {(2,10)/z3/$z_3$},
              {(1,9)/z4/$z_4$}, {(2,9)/z5/$z_5$}, {(3,9)/z6/$z_6$}} %
        \node[vertex] (\name) at \pos {\label};
    \foreach \source / \dest in {z1/z2, z2/z3, z4/z3, z4/z5, z5/z6}
        \path[edge] (\source) -- (\dest);
        
    \foreach \pos / \name / \label in {{(5,9.5)/z21/$z_1$}, {(6,9.5)/z22/$z_2$}, {(7,9.5)/z23/$z_3$}}
        \node[vertex] (\name) at \pos {\label};
    \foreach \source / \dest in {z21/z22, z22/z23}
        \path[edge] (\source) -- (\dest);
        
     \node at (-3.5,8) {$O_{111}$};
     \node at (-1,8) {$\containedIn$};
     \node at (1.5,8) {$O_{11011}$};
     \node at (3.75,8) {$\containedIn$};
     \node at (6,8) {$O_{11}$};

\end{tikzpicture}
\caption{Oriented path queries $O_{111}$, $O_{11011}$ and $O_{11}$.}
\label{fig:OPQ-containment}
\end{figure*}

Only at length $5$ does one encounter an oriented path query that is not equivalent
to a path query, namely $P_{11011}$, shown in Figure~\ref{fig:OPQ-containment}.  Note that,
although there is a homomorphism from $O_{11}$ to $O_{11011}$
and from $O_{11011}$ to $O_{111}$, there is no homomorphism 
from $O_{11011}$ to $O_{11}$ or from $O_{111}$ to $O_{11011}$.
So $O_{11011}$ is ``between'' a path query of length $2$ and one of length $3$.
How many oriented path queries can appear between two ``consecutive'' path queries?
Unfortunately, it seems that there can be an unbounded number of 
oriented path queries between two consecutive path queries,

\begin{theorem}
There exist (oriented) path queries $P_{i}$ and $P_{i+1}$ for which there is no oriented
path query $O$ such that $P_{i} \containedIn O \maxContainedIn P_{i+1}$.
\end{theorem}

\begin{proof}
Consider path queries $P_3 = O_{111}$ and $P_2 = O_{11}$, as well as
the oriented path query $O_{11011}$ shown in Figure~\ref{fig:OPQ-containment}.
We can produce any number of OPQs by ``iterating'' a sub-pattern in $O_{11011}$.
Let $Q_0 = O_{111}$ and let $Q_i = O_{11(01)^i1}$, $i \geq 1$,
as shown in Figure~\ref{fig:OPQ-template}.
A homomorphism from $Q_{i+1}$ to $Q_i$ is given by mapping 
$w_{2i+1}$ and $w_{2(i+1)}$ in $Q_{i+1}$ to 
$w_{2j-1}$ and $w_{2j}$ in $Q_i$, respectively 
(or $w_1$ and $w_2$ to $z_2$ and $z_3$, respectively, for $Q_1$ to $Q_0$).
There is a homomorphism from $O_{11}$ to each $Q_i$, since
each $Q_i$ contains $O_{11}$ as a sub-path.
So we have the following:
$$
O_{111} \containedIn Q_1 \containedIn \cdots \containedIn Q_i \containedIn \cdots \containedIn O_{11}
$$
We claim that none of the reverse containments to those above hold.

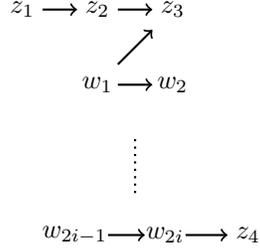
\begin{figure}\centering
\begin{tikzpicture}[auto, swap]
\tikzstyle{vertex}=[minimum size=15pt,inner sep=0pt]
\tikzstyle{edge} = [draw,thick,->]
\tikzstyle{dot} = [draw,thick,dotted,-]

    \foreach \pos / \name / \label in {{(0,10)/z1/$z_1$}, {(1,10)/z2/$z_2$}, {(2,10)/z3/$z_3$},
              {(1,9)/w1/$w_1$}, {(2,9)/w2/$w_2$}, {(0.7,7)/wi/$w_{2i-1}$}, {(1.9,7)/w2i/$w_{2i}$}, {(3,7)/z4/$z_4$}} %
        \node[vertex] (\name) at \pos {\label};
    \foreach \source / \dest in {z1/z2, z2/z3, w1/z3, w1/w2, wi/w2i, w2i/z4}
        \path[edge] (\source) -- (\dest);
        
   \node (d1) at (1.5,8.4) {};
   \node (d2) at (1.5,7.4) {};
   \path[dot] (d1) -- (d2);

\end{tikzpicture}
\caption{Oriented path query $Q_i = O_{11(01)^i1}$.}
\label{fig:OPQ-template}
\end{figure}

There can be no homomorphism from $O_{111}$ to $Q_1$ because
the longest directed path in $Q_1$ is of length only $2$.  Similarly,
there can be no homomorphism from $Q_i$ to $O_{11}$ because
one of the $O_{11}$ components of $Q_i$ (from $z_1$ to $z_3$,
or from $w_{2i-1}$ to $w_{2i}$) must map
to the directed path of length $2$ in $O_{11}$, but then either 
$z_1$ or $z_4$ cannot be mapped to any variable in $O_{11}$.

Now consider a possible homomorphism from $Q_i$ to $Q_{i+1}$.
Variables $z_1$, $z_2$ and $z_3$ in $Q_i$ could map to variables
$w_{2i+1}$, $w_{2(i+1)}$ and $z_4$ in $Q_{i+1}$, respectively,
but, once again, $z_4$ in $Q_i$ cannot be mapped anywhere.
So $z_1$, $z_2$ and $z_3$ in $Q_i$ must map to $z_1$, $z_2$ 
and $z_3$, respectively, in $Q_{i+1}$.
Now, whether the $w$ variables in $Q_i$ map to distinct $w$ variables
in $Q_{i+1}$ or not, $z_4$ in $Q_i$ cannot be mapped to $Q_{i+1}$
because $Q_{i+1}$ has one more $01$ pattern than $Q_i$.

Perhaps there is some other query $R$ that is contained in $O_{11}$
and contains each $Q_i$.  
Because $R$ is of finite size, no homomorphism from $R$ can map
one variable to one of $z_1$ to $z_3$ and another variable to $z_4$
(since $Q_i$ can be ``pumped'' to $Q_j$, $j > i$).  Hence, variables
in $R$ must map to some subset of $\{z_1, z_2, z_3, w_1, \ldots, w_k \}$,
or to some subset of $\{ w_{2i-k}, \ldots, w_{2i}, z_4 \}$, for some fixed $k$.
But the queries induced by these subsets of variables are all equivalent
to $O_{11}$.  Hence, $R$ (which is always assumed to be minimal) must
have fewer variables than $O_{11}$.
The only queries contained in $O_{11}$
having fewer variables than $O_{11}$ are $C_1$ and $C_2$, 
the cycles with one and two variables, respectively.
But neither of these is an oriented path query (nor contains any $Q_i$).
We conclude that there is no query ``between'' $O_{111}$ and
$O_{11}$ (including $O_{111}$ itself) that is maximally contained in $O_{11}$.
\end{proof}

The above theorem shows that the partial order $(\mathcal{S}, \containedIn)$, where $\mathcal{S}$ is
the set of oriented path queries, does not satisfy the ascending chain
condition.
A partial order $(\mathcal{S}, \containedIn)$ satisfies the {\em ascending chain condition}
if no infinite strictly ascending sequence
$$
q_1 \propContainedIn q_2 \propContainedIn q_3 \propContainedIn \cdots
$$
of elements of $\mathcal{S}$ exists.

Note that, because of the equivalences shown in Figure~\ref{fig:OPQ-equivalence},
we can reduce the problem of determining maximal containment for 
oriented path queries to that of path queries if we limit the number
of occurrences of the same relation name in an oriented path query to four.

\noop{
In the next section, we consider the problem of deciding maximal containment,
i.e., given CQs $Q_1$ and $Q_2$, is $\query_1 \maxContainedIn \query_2$?
If we do not restrict the numbers of occurrences of relation names in the CQs,
we will fall foul of the above result.  Hence,  we restrict our attention to CQs 
in which each relation name can appear at most twice in the body.
Another reason for limiting the number of occurrences to two is that Saraiya~\cite{Saraiya90} 
shows that $Q_1 \containedIn Q_2$ can be tested in PTIME if each atom in $Q_2$ can be mapped 
to at most two atoms in $Q_1$.  Increasing that number to three 
results in NP-completeness.

In addition, many queries in practice refer to the same relation at most twice.
For example, as an important practical class of queries, every {\em star} query is (semantically) a 2CQ.  Here a star query is such that (1) all body atoms are binary and (2) there exists exactly one variable $x$ which appears exactly once in each atom.  Note that relation names can appear any number of times in the body; it is easy to establish that the core of every star query is a 2CQ.
It has been observed in recent large-scale query log analyses that typically over 90\% of queries in practice are star queries \cite{BonifatiMT19,BonifatiMT20}.
}
\section{Maximal containment of 2CQs}
We next consider the problem of deciding maximal containment,
i.e., given CQs $Q_1$ and $Q_2$, is $\query_1 \maxContainedIn \query_2$?
If we do not restrict the numbers of occurrences of relation names in the CQs,
we will fall foul of the result established in the previous section.  Hence,  we restrict our attention to CQs 
in which each relation name can appear at most twice in the body.
Another reason for limiting the number of occurrences to \emph{two} is that Saraiya~\cite{Saraiya90} 
shows that $Q_1 \containedIn Q_2$ can be tested in PTIME if each atom in $Q_2$ can be mapped 
to at most \emph{two} atoms in $Q_1$.  Increasing that number to three 
results in NP-completeness.

In addition, many queries in practice refer to the same relation at most twice.
For example, as an important practical class of queries, every {\em star} query is (semantically) a 2CQ.  Here a star query is such that (1) all body atoms are binary and (2) there exists exactly one variable $x$ which appears exactly once in each atom.  Note that relation names can appear any number of times in the body; it is easy to establish that the core of every star query is a 2CQ.
It has been observed in recent large-scale query log analyses that typically over 90\% of queries in practice are star queries \cite{BonifatiMT19,BonifatiMT20}.

\subsection{2CQs and their restrictions}
A conjunctive query in which each relation name appears at most twice
(2CQ) is a rule of the form
\begin{eqnarray*}
    ( z_1, \ldots, z_m ) & \gets & S_1(\overline{x}_1), \ldots, S_n(\overline{x}_n)
\end{eqnarray*}
where $S_i = S_j$ for at most two indexes $i$ and $j$.

For the class of 2CQs we have a characterisation of maximal
containment based on the notion of a {\em restriction} of a query.

\begin{definition} \label{def:restriction}
Let $\query = (\overline{z}) \gets  S_1(\overline{x}_1), \ldots, S_n(\overline{x}_n)$
be a 2CQ query, and $vars(\query)$ denote the set of variables appearing in \query.
A {\em restriction} of $\query$ takes one of the following four forms:
\begin{enumerate}
\item $h(\overline{z}) \gets S_1(h(\overline{x}_1)), \ldots, S_n(h(\overline{x}_n))$,
where $h$ is a function from $vars(\query)$ to $vars(\query)$ 
which is the identity on all variables in $vars(\query)$ except for one variable $y$,
such that $h(y)$ is distinguished if $y$ is.

\item $(\overline{z}) \gets S_1(\overline{x}_1), \ldots, S_n(\overline{x}_n), T(\overline{y})$,
where $T$ is a relation name in $\sigma$ distinct from those in $Q$,
and $\overline{y}$ consists of distinct new variables not appearing in $vars(Q)$.

\item $(\overline{z}) \gets S_1(\overline{x}_1), \ldots, S_n(\overline{x}_n), T(h(\overline{y}))$,
where 
\begin{itemize}
\item $T = S_j$ for exactly one $j \in [1..n]$,
\item $\overline{y}$ consists of distinct new variables not appearing in $vars(Q)$,
\item $h$ is a function from $Y$ to $vars(\query) \cup Y$ 
(where $Y$ is the set of variables appearing in $\overline{y}$) 
which is the identity on all variables in $Y$ except for exactly one variable.
\end{itemize}

\item $(\overline{z}) \gets S_1(\overline{x}_1), \ldots, S_n(\overline{x}_n), T_1(h(\overline{y}_1)), T_2(\overline{y}_2)$,
where 
\begin{itemize}
\item $T_1 = S_j$ and $T_2 = S_k$ ($S_j \neq S_k$) for exactly one $j \in [1..n]$ and exactly one $k \in [1..n]$, 
\item $\overline{y}_1$ and $\overline{y}_2$ are tuples of distinct variables not in $vars(\query$), and 
\item $h$ is a function from $Y_1$ to $Y_1 \cup Y_2$ 
(where $Y_1$ and $Y_2$ are the sets of variables appearing in $\overline{y}_1$ and $\overline{y}_2$, respectively) 
which is the identity on all variables in $Y_1$ except for exactly one variable which $h$ maps to a variable in $Y_2$.
\end{itemize}
\end{enumerate}
We refer to a restriction corresponding to case~(i) above
as a {\em type-$i$ restriction}.
\end{definition}

The following proposition clearly holds by virtue of the above definition.

\begin{proposition}
Each restriction of a 2CQ is a 2CQ.
\end{proposition}

However, a restriction of a minimal 2CQ does not necessarily result in a 
minimal 2CQ, except for restrictions of type~2 which are always minimal.
Also, there can be containment relationships among
the various restrictions of a query, as shown in the following example.

\begin{example}
Consider the following queries:
\begin{eqnarray*}
Q_1: \ \  () & \leftarrow & R(x, y) \\
Q_2: \ \  () & \leftarrow & R(x, y), R(y, z) \\
Q_3: \ \  () & \leftarrow & R(x, y), R(y, x) \\
Q_4: \ \  () & \leftarrow & R(x, x)
\end{eqnarray*}
We have $Q_4 \containedIn Q_3 \containedIn Q_2 \containedIn Q_1$.
A type-$1$ restriction can map $y$ to $x$, producing
$Q_4$ directly from $Q_1$, while there is a type-$3$ restriction of $Q_1$,
namely $Q_2$, which contains $Q_4$.  Considering $Q_2$, the type-$1$
restriction mapping $z$ to $x$ produces $Q_3$, which is maximally contained
in $Q_2$.  However, mapping $z$ to $y$, $y$ to $z$, $y$ to $x$ or $x$ to $y$
would each produce $Q_4$ directly, which is obviously not maximally contained
in $Q_2$.
\end{example}

The above examples show that there can be containment relationships
among the queries generated by type-$1$, $3$ and $4$ restrictions;
hence, we need to perform a containment check when generating them.  
This check is not necessary for type-$2$ restrictions as will be shown below.

Let $Q$ be a minimal 2CQ and $QR$ be the set of type $1$, $3$ or $4$ restrictions
applied to $Q$.  The {\em reduced set of type $1$, $3$ or $4$ restrictions} of $Q$
is given by
$$ \{ Q'  \mid  Q' \mbox{ is the core of a query in } QR, Q \not\containedIn Q', \mbox{ and} 
\not\exists Q'' \in QR \: (Q'' \neq Q') \mbox{ such that } Q' \containedIn Q'' \} . 
$$

We say that queries $Q_1$ and $Q_2$ are {\em incomparable} if
$Q_1 \not\containedIn Q_2$ and $Q_2 \not\containedIn Q_1$.

\begin{lemma}\label{lem:type-2}
Let $Q$ be a minimal 2CQ, $Q'$ be a type-$2$ restrictions of $Q$,
and $Q''$ be either of type-$2$ (in which case it is assumed to be
non-isomorphic to $Q'$) or in the reduced set of type $1$, $3$ or $4$ restrictions of $Q$.
Then (1)~$Q'$ is minimal, (2)~there is no homomorphism from $Q'$ to $Q$, and
(3)~$Q'$ and $Q''$ are incomparable.
\end{lemma}

\begin{proof}
(1) Let $A$ be the atom added to $Q$ to form $Q'$,
and $R$ be its relation name.  Since $Q$ is minimal
and $R$ does not appear in $Q$, there can be no
homomorphism from $Q'$ to $Q'$ in which one atom
maps to another.  This shows both that $Q'$ is minimal 
and that there is no homomorphism from $Q'$ to $Q$.

Next we show that $Q'$ and $Q''$ are incomparable.
Assume first that $Q''$ is a type-$2$ restriction.
There can be no homomorphism between
$Q'$ and $Q''$ since they each contain an atom 
with a relation name not appearing in the other.

Now assume that $Q''$ is of type $1$, $3$ or $4$.
There can be no homomorphism from $Q'$ to $Q''$
because $Q'$ contains an atom with a relation name 
not appearing in $Q''$.   
There can be no homomorphism $g$ from $Q''$
to $Q'$ because, in each case, $g$ 
would have to map the atoms of $Q''$ to $S_1(\overline{x}_1), \ldots, S_n(\overline{x}_n)$,
and therefore $Q'' \equiv Q$ which contradicts the fact that $Q''$
is in the reduced set of type $1$, $3$ or $4$ restrictions of $Q$.
\end{proof}

The {\em reduced set of restrictions} of $Q$, denoted $RR(Q)$, is the union of the set
of type-$2$ restrictions of $Q$ and the reduced set of type $1$, $3$ or $4$ 
restrictions of $Q$.

\begin{proposition}\label{prop:2CQ-max}
If $R$ and $Q$ are minimal 2CQs, then $R \maxContainedIn Q$
if and only if $R \in RR(Q)$.
\end{proposition}

\begin{proof}
Let $M(Q)$ be the set of cores of queries maximally contained in 2CQ $Q$.
An alternative definition of $M(Q)$ is as the set of incomparable cores of queries
properly contained in $Q$ such that for each 2CQ $P \propContainedIn Q$,
there exists an $R \in M(Q)$ such that $P \containedIn R$.
Below we show that $RR(Q)$ is equal to $M(Q)$ in three steps:
(1)~if $R \in RR(Q)$, then $R \propContainedIn Q$,
(2)~if $R \in M(Q)$, then $R \in RR(Q)$, and 
(3)~if $R \in RR(Q)$, then there is no $R' \in RR(Q)$ ($R' \neq R$)
such that $R \propContainedIn R''$.

Steps (1) and (3) are straightforward, so we deal with them first.
For (1), let $Q$ be a 2CQ and $R \in RR(Q)$.
If $R$ is a type-1 restriction, then the function $h$ in the definition
of a type-$1$ restriction provides a homomorphism from $Q$ to $R$.
If $R$ is a type-2, type-3 or type-4 restriction, then the identity function on
$vars(Q)$ provides a homomorphism from $Q$ to $R$.
Thus, $R \containedIn Q$.
Proper containment follows from the definition of $RR(Q)$ and Lemma~\ref{lem:type-2}.

For step (3), the definition of $RR(Q)$, along with Lemma~\ref{lem:type-2},
ensures that if $R \in RR(Q)$, then there does not exist another restriction 
$R'$ of $Q$ ($R' \neq R$) such that $R \containedIn R'$. 

We now consider step (2), namely showing that every query maximally contained
in 2CQ $Q$ is in the reduced set of restrictions of $Q$.  
Let $Q$ be the 2CQ query $(\overline{z}) \gets S_1(\overline{x}_1), \ldots, S_n(\overline{x}_n)$.
We use the fact that $M(Q)$ is the set of incomparable cores of queries
properly contained in $Q$ such that for each 2CQ $P \propContainedIn Q$,
there exists an $R \in M(Q)$ such that $P \containedIn R$.
We show that, for each 2CQ $P$ such that $P \propContainedIn Q$,
there is an $R \in RR(Q)$ such that $P \containedIn R$,
and hence that $R \in M(Q)$.  

Choose some 2CQ $P \propContainedIn Q$.
Since $P \propContainedIn Q$, there is some homomorphism $g$ from $Q$ to $P$.
Let us call the atoms in $P$ to which the atoms of $Q$ are mapped by $g$
the {\em target\/} atoms, and any additional atoms in $P$ the {\em non-target\/}
atoms.  We will also call two atoms with the same relation name {\em counterparts}.

For the homomorphism $g$ to exist, $P$ must include at least one atom for each
relation name $S_i$ in $Q$, as well as possibly atoms with other relation names.
Also, the head of $P$ must be $g(\overline{z})$.

Assume first that $P$ contains an atom $T(\overline{w})$ where $T$
is different from each of $S_1, \ldots, S_n$.  
Let $R$ be the type-2 restriction of $Q$ that includes 
$T(\overline{y})$ as its additional atom.  Since $R$
is identical to $Q$ on its $S_i$ atoms, $g$ maps
the $S_i$ atoms of $R$ to atoms in $P$ as well.  
Then $g$ can be extended to homomorphism $g'$ 
which also maps $T(\overline{y})$ in $R$ to 
$T(\overline{w})$ in $P$, since the variables in $\overline{y}$ 
are distinct and disjoint from those in the rest of $R$.  
We conclude that $P \containedIn R$.

Now assume that $P$ includes only atoms using relation names $S_1$ to $S_n$.
There are two cases to consider: (a)~no relation name appears more often
in $P$ than it does in $Q$, and (b)~some relation name appears
more often in $P$ than it does in $Q$.

Consider case (a).  
It cannot be the case that each variable in $Q$ is mapped by $g$ to a distinct
variable in $P$, for otherwise $g$ would be a bijection, i.e., simply a renaming 
of variables, every atom of $P$ would be a target, and $P \equiv Q$.
So let $u$ and $v$ be two distinct variables in $Q$ that are mapped by $g$
to the same variable $w$ in $P$.  Without loss of generality, assume that,
if one of $u$ or $v$ is distinguished, then it is $v$ which is distinguished.
Let $R$ be the type-$1$ restriction 
of $Q$ in which $u$ is the variable on which function $h$ is not
the identity, and such that $h$ maps $u$ to $v$.  
Let $g'$ be the homomorphism $g$ excluding
the mapping for $u$.  We claim that $g = g' \circ h$ and
hence that $g'$ is a homomorphism from $R$ to $P$.
Let $y$ be a variable in $Q$.  If $y \neq u$, then $h$
is the identity on $y$ and, by definition, $g'(y) = g(y)$.
If $y=u$, then $h(y)=v$, and, once again, $g'(y) = g(y)$ since $g'(v)=g(v)$.
We conclude that $g'$ is a homomorphism from $R$ to $P$,
and hence that $P \containedIn R$.

Now consider case (b), i.e.\ $P$ contains only atoms using relation names
$S_1$ to $S_n$ and at least one relation name appears more often
(i.e., twice) in $P$ than it does in $Q$ (i.e, once).  
So there is at least one non-target atom in $P$.  

We first show that (at least) one of the following three conditions must be true in $P$:
\begin{enumerate}\renewcommand{\labelenumi}{(\roman{enumi})}
\item There is a non-target atom which shares 
a variable with a target atom. 

\item There is a non-target atom in which the same
variable appears in two different positions.

\item There is a pair of non-target atoms which share a variable. 
\end{enumerate}

Assume to the contrary that none of the above three conditions is true.
We know there is at least one non-target atom $T(\overline{z})$.
By assuming~(i), (ii) and~(iii) above to be false, the $z_i$ must be
distinct and appear nowhere else in $P$.
Furthermore, no $z_i$ can be distinguished, since that would
require $z_i$ also to appear in a target atom.
Hence, there is a homomorphism which maps
$T(\overline{z})$ to its target counterpart, meaning $P$ is not minimal, 
a contradiction.
We conclude that $P$ must satisfy at least one of the conditions~(i),
(ii) or~(iii) above.  We consider each in turn below.

Assume that $P$ satisfies condition~(i), and let $T(\overline{z})$
be the non-target atom which satisfies the condition.  Assume that
$T(\overline{z})$ shares variable $z_j$ with target atom $S_k(\overline{w}_k)$ in $P$.
Let $S_i(\overline{x}_i)$ be the atom in $Q$ mapped to $S_k(\overline{w}_k)$ in $P$
by homomorphism $g$, with $x_{ij}$ being the variable mapped to $z_j$.
Now let $R$ be the type-3 restriction of $Q$ which adds $T(\overline{y})$
to $Q$ and maps $y_j$ to $x_{ij}$.  
Clearly $g$ can be extended to a homomorphism $g'$ from
$R$ to $P$ by including mappings for each of the distinct variables
in $T(\overline{y})$ not shared with any atom to their corresponding
variables in $T(\overline{z})$ in $P$, while the shared variable $x_{ij}$
in $T(\overline{y})$ in $R$ is mapped to the correct shared variable 
$z_j$ in $P$.

Now assume that $P$ satisfies condition~(ii) and let $T(\overline{z})$
be the non-target atom such that $z_j = z_k$.
Let $R$ be the type-3 restriction of $Q$ which adds $T(\overline{y})$
to $Q$ and $h$ maps $y_j$ to $y_k$.  
Since $T(\overline{y})$ in $R$ shares no variables with other atoms in $R$, 
the homomorphism $g$ from $Q$ to $P$ can be extended to a homomorphism $g'$ 
from $R$ to $P$ by mapping the variables in $T(\overline{y})$,
including the shared variable, to those in $T(\overline{z})$. 

Finally assume that $P$ satisfies condition~(iii) and let $T_1(\overline{u})$
and $T_2(\overline{v})$ be the two non-target atoms such that
$u_i = v_j$.  Let $R$ be the type-$4$ restriction of $Q$ which
adds $T_1(h(\overline{y}_1))$ and $T_2(\overline{y}_2)$ to $Q$
and maps $y_{1i}$ to $y_{2j}$.  Since the $T_i$ atoms in $R$ share 
no variables with other atoms in $R$, the homomorphism $g$ 
from $Q$ to $P$ can be extended to a homomorphism $g'$ 
from $R$ to $P$ which maps $T_1(\overline{y}_1)$ to $T_1(\overline{u})$
and $T_2(\overline{y}_2)$ to $T_2(\overline{v})$. 

We conclude that, for each minimal 2CQ $P$ such that $P \propContainedIn Q$,
there is a restriction $R \in RR(Q)$ such that 
$P \containedIn R$, and hence that $R \in M(Q)$. 
\end{proof}

\subsection{Complexity of maximal containment}
Based on Saraiya's result~\cite{Saraiya90}, we next show that maximal containment
can be decided in polynomial time.  Let $\sigma$ be a relational schema
and let $\langTwoCQ$ denote the set of all 2CQs of arity $\alpha$ defined on $\sigma$.
Given a query $\query \in \langTwoCQ$
\begin{eqnarray*}
    ( z_1, \ldots, z_{\alpha} ) & \gets & S_1(\overline{x}_1), \ldots, S_n(\overline{x}_n)
\end{eqnarray*}
its {\em size} is given by $\alpha + n + \sum_{i=1}^n |\overline{x}_i|$.

\begin{proposition}
Given relational schema $\sigma$ and queries $\query_1, \query_2 \in \langTwoCQ$,
deciding $\query_1 \maxContainedIn \query_2$ can be done in time polynomial in
the sizes of $\query_1$ and $\query_2$.
\end{proposition}

\begin{proof}
From~\cite{Saraiya90} we have that $\query_1 \containedIn \query_2$ and
$\query_1 \equiv \query_2$ can be decided in polynomial time.  
Minimising a 2CQ $\query$ can also be achieved in polynomial time
by simply removing each atom from the body of $\query$ in turn and checking for containment.
After minimising both $\query_1$ and $\query_2$, we can use Proposition~\ref{prop:2CQ-max}
to decide maximal containment by generating all minimal 2CQ restrictions of $\query_2$,
testing each for equivalence to $\query_1$.  We need to show that there are only
a polynomial number of such restrictions.

Assume $\query_2$ contains $\alpha$ head variables and $n$ body atoms whose total arity is $s$.
Assume $\query_1$ uses $k$ relation names in its body atoms.  
For type-$1$ restrictions, $\query_2$ has at most $s$ variables,
so there are at most $s(s-1)/2$ type-$1$ restrictions.
Query $\query_2$ has at most $k$ type-$2$ restrictions that could possibly be
equivalent to $\query_1$.  
Each type-$3$ restriction introduces an atom whose relation name appears in $\query_2$,
with either one variable shared 
with an atom in $\query_2$ or two variables the same.  The former case generates
at most $s^2$ restrictions, while the latter generates
at most $s (s-1)/2$ restrictions.
Type-$4$ restrictions equate pairs of variables in two atoms whose relation names
appear in $\query_2$, so there can be at most $s(s-1)/2$ type-$4$ restrictions.
Thus, the total number of restrictions of $\query_2$ which have to be checked
for equivalence to $\query_1$ is polynomial in the sizes of $\query_1$ and $\query_2$.
\end{proof}

\section{A semantic metric for 2CQs}

\subsection{Well-defined and well-behaved containment-based distance}

We are ready to define our distance function on 2CQs of arity $\alpha$ over schema $\sigma$.

\begin{definition} \label{def:distanceContainment}
    For $\query_1, \query_2 \in \langTwoCQ$, let $\distCont:\langTwoCQ \times \langTwoCQ \to [0, \infty)$ be defined as follows.
    \begin{itemize}
        \item $\distCont(\query_1, \query_2) = 0$ if and only if $\query_1 \equiv \query_2$.
        \item $\distCont(\query_1, \query_2) = 1$ if and only if
            $\query_1 \maxContainedIn \query_2$ or $\query_2 \maxContainedIn \query_1$. 

        \item $\distCont(\query_1, \query_2) = i>1$ if and only if
            (a) there exists $\query\in \langTwoCQ$ such that $\distCont(\query_1, \query)=i-1$ and $\distCont(\query_2,\query) =1$; and
            (b) there does not exist $\query'\in\langTwoCQ$ such that
            $\distCont(\query_1, \query')=i'$, $\distCont(\query_2,\query') =1$, and $i'< i-1$. 
    \end{itemize}
\end{definition}

\begin{example}
Consider again the 2CQs $Q_1$:
\begin{eqnarray*}
(x, y) & \leftarrow & R(x, y), R(y, x), L(x), L(y)
\end{eqnarray*}
and $Q_2$:
\begin{eqnarray*}
(x, y) & \leftarrow & R(x, z), L(y), L(z)
\end{eqnarray*}
from Example~\ref{ex:homomorphism}.
There are 2CQs $Q_3$
\begin{eqnarray*}
(x, y) & \leftarrow & R(x, x), L(x), L(y)
\end{eqnarray*}
and $Q_4$:
\begin{eqnarray*}
(x, x) & \leftarrow & R(x, x), L(x)
\end{eqnarray*}
such that $Q_4 \maxContainedIn Q_3 \maxContainedIn Q_2$ and $Q_4 \maxContainedIn Q_1$.
Hence, $\distCont(\query_1, \query_4) = 1$, $\distCont(\query_2, \query_3) = 1$ and
$\distCont(\query_3, \query_4) = 1$.
So we have that $\distCont(\query_2, \query_4) = 2$, and since it turns out that there is 
no query $Q'$ such that $\distCont(\query_1, \query') = 1$ and $\distCont(\query_2, \query') = 1$,
we have that $\distCont(\query_1, \query_2) = 3$.
\end{example}

We establish next that \distCont{} is well-defined.
\begin{theorem}\label{thm:distCont-well-defined}
For every $\query_1, \query_2\in \langTwoCQ$, there exists a unique $i\in [0, \infty)$ such that 
$\distCont(\query_1, \query_2) = i$.
\end{theorem}

\begin{proof}
If $\query_1 \equiv_{\pi} \query_2$, then by the first case of Definition~\ref{def:distanceContainment},
we have that $\distCont(\query_1, \query_2) = i=0$.  As the second and third
cases of the Definition require $\query_1 \not\equiv_{\pi} \query_2$, it also cannot be that 
there exists $j>0$ such that $\distCont(\query_1, \query_2) =j$.
Hence, $i$ is unique.  

Now assume that $\query_1 \not\equiv_{\pi} \query_2$ so $\distCont(\query_1, \query_2) > 0$.
Also assume by way of contradiction that $\distCont(\query_1, \query_2)$ is not unique, 
i.e., $\distCont(\query_1, \query_2) = i$ and $\distCont(\query_1, \query_2) = j$,
for some $i$ and $j$, $i \neq j$.  Without loss of generality, assume that $i < j$.
If $\distCont(\query_1, \query_2)$ is $j$ ($i$), then by condition~(a) in the third case 
of the definition, there is a query $Q$ ($Q'$) such that $\distCont(\query_2, \query) = 1$
($\distCont(\query_2, \query') = 1$) and $\distCont(\query_1, \query) = j-1$ 
($\distCont(\query_1, \query') = i-1$).
But then $Q'$ violates condition~(b) in the definition for distance $j$ since 
($\distCont(\query_2, \query') = 1$) and $\distCont(\query_1, \query') = i-1 < j-1$.
We conclude that $\distCont(\query_1, \query_2)$ is unique.
\end{proof}

We also establish that \distCont{} is well-behaved.
In particular, we show that \distCont{} is a {\em metric}, i.e., is positive
and symmetric, preserves the identity of indiscernibles, and satisfies the
triangle inequality.  

\begin{definition} \label{def:metric}
    A binary function $f:S^2 \to [0, \infty)$
    is a {\em metric} on $S$ if and only if for all $x, y, z \in S$ it holds that
    \begin{enumerate}
        \item $f(x, y) \geq 0$;
        \item $f(x, y) = f(y, x)$; 
        \item $f(x, x) = 0$; and
        \item $f(x, y) \leq f(x, z) + f(z, y)$.
    \end{enumerate}
\end{definition} 
\noindent Being a metric is a crucial property for applications of
distance functions~\cite{ChenGZJYY17,ZezulaADB06}.

\begin{theorem}\label{thm:distCont-is-metric}
    \distCont{} is a metric on \langTwoCQ.
\end{theorem}

\begin{proof}
Properties (1), (2) and (3) follow straightforwardly from Definition~\ref{def:distanceContainment}.
The proof of property~(4) proceeds by induction on the distance between two 2CQs.
The base case when the distance is zero is obvious.  Our inductive hypothesis is that, 
for all 2CQs $\query_1$, $\query_2$ and $\query_3$ such that 
$\distCont(\query_1, \query_3) \leq k$, it is the case that 
$\distCont(\query_1, \query_3) \leq \distCont(\query_1, \query_2) + \distCont(\query_2, \query_3)$.

Now let $\query_1$ and $\query_3$ be 2CQs such that $\distCont(\query_1, \query_3)= k+1$.
From Definition~\ref{def:distanceContainment}, we know that there exists a 2CQ $\query_2$
such that $\distCont(\query_1, \query_2) = k$ and $\distCont(\query_3, \query_2) = 1$,
and no other query at distance one from $\query_3$ which has distance less than $k$ from $\query_1$.
Consider an arbitrary 2CQ $\query'$.
By the inductive hypothesis, 
$\distCont(\query_1, \query_2) \leq \distCont(\query_1, \query') + \distCont(\query', \query_2)$.
Since $\distCont(\query_3, \query_2) = 1$, we have 
$\distCont(\query_1, \query_3) = k+1 \leq \distCont(\query_1, \query') + \distCont(\query', \query_2) + 1$.
In other words, 
$\distCont(\query_1, \query_3) \leq \distCont(\query_1, \query') + \distCont(\query', \query_3)$,
as required.
\end{proof}

\subsection{Complexity of computing semantic distance}
In order to compute the distance between a pair of 2CQs of the same arity over the same schema,
we can construct a graph based on the maximal containment relationship between queries.  
Given schema $\sigma$ and arity $\alpha$, the
{\em maximal containment graph\/} ({\em MC-graph\/} for short) $\graph =(V,E)$ 
for $\sigma$ and $\alpha$ is constructed as follows: 
the set of nodes $V$ comprises the set of minimal 2CQs of arity $\alpha$
that can be constructed for $\sigma$; there is a directed edge $e = (u,v)$ in $E$ if and only if
$u$ maximally contains $v$.

We view $\graph$ being orientated such that edges are directed vertically downwards.
Then a {\em top} node (query) in $\graph$ is a node $v$ such that there is no node $u$
with $(u,v) \in E$.  A {\em bottom} node (query) in $\graph$ is a node $u$ such that 
there is no node $v$ with $(u,v) \in E$.

It is not hard to see that there is a unique bottom node (up to renaming of variables)
in $\graph$, corresponding to the 2CQ
\begin{eqnarray*}
    (\overline{y}) & \gets & S_1(\overline{x}_1), \ldots, S_n(\overline{x}_n)
\end{eqnarray*}
where $\sigma = \{ S_1, \ldots, S_n \}$ and there is a variable $x$ such that 
$\overline{y}$ is a tuple of $\alpha$ copies of $x$ and each $\overline{x}_i$ is 
a tuple of $arity(S_i)$ copies of $x$.
There are, however, an exponential number of top nodes (queries) in $\graph$.
Each top query has a distinct variable in each position of each atom.  Given a set of 
relation names $\rho$, we define the set of {\em top queries associated with\/} $\rho$,
denoted $tq(\rho)$, to be the cores of those top queries whose body comprises two atoms for each 
relation name in $\rho$, and whose head consists of a permutation of $\alpha$ variables
from the body.  If there are fewer than $\alpha$ variables in the body, then some
variables in the head are repeated in order to extend it to $\alpha$ variables.
In $\graph$, there is a set of top queries $tq(\sigma')$ for each non-empty 
$\sigma' \subseteq \sigma$.

\begin{theorem}\label{th:distance-complexity}
Given schema $\sigma$ and $\query_1, \query_2 \in \langTwoCQ$, $\distCont(\query_1, \query_2)$ 
can be computed in exponential time.
\end{theorem}

\begin{proof} 
We can construct $\graph = (V,E)$ as follows.  Initially $V$ is given by $tq(\sigma')$ 
for each non-empty $\sigma' \subseteq \sigma$.  Next, we successively add new nodes to $V$ 
by finding the reduced set of restrictions of each node in $V$, until no new 
(i.e., non-equivalent) nodes can be found.
Although we start from a number of nodes which is exponential in the size of $\sigma$
and $\alpha$, the size of $\graph$ remains exponential, although in $\alpha$ plus
the total arity of all relation names in $\sigma$.

Now given $\query_1, \query_2 \in \langTwoCQ$, we can search $\graph$ to find a 2CQ
equivalent to $\query_1$, say, in exponential time, and then use a shortest-path algorithm 
(viewing the edges in $\graph$ as undirected)
to find $\distCont(\query_1, \query_2)$, once again in exponential time.
\end{proof}

\section{Related Work}

Refinement operators have been used in inductive logic programming for concept learning 
(model inference)~\cite{Vanderlaag98}.  A refinement operator computes a specialisation
or generalisation of a clause, based on the notion of subsumption between clauses.
The downward refinement operator defined in~\cite{Vanderlaag98} is analogous to our restriction operator,
although it includes only rules corresponding to some of our types of 
restriction (as well as one for introducing function symbols).
In order to ensure completeness (i.e., that their operator
returns all specialised clauses), the authors of~\cite{Vanderlaag98} need to include all non-reduced clauses
(cf.\ non-minimal CQs) equivalent to a given clause in their search space.
We avoid this by including other types of 
restriction which ensure completeness when restricted to 2CQs.

More recently, refinement operators have been applied to CQs in description logic (DL)
in order to measure the similarity between individuals~\cite{Sanchez-Ruiz16}.
The downward operator of~\cite{Sanchez-Ruiz16}, defined using several rewriting rules, 
produces from a DL CQ $Q$ all queries which are properly contained in $Q$.
However, a rule is not guaranteed to produce a 
query which is {\em maximally} contained in $Q$, unlike in our approach.

In work studying which classes of conjunctive queries are uniquely
characterised by polynomially many positive and negative examples~\cite{Cate22},
ten Cate and Dalmau define the {\em frontier} of a finite structure
(and conjunctive query) in the homomorphism lattice of such structures
(queries).  In our setting, the frontier of a conjunctive query $Q$ 
is essentially the set of queries each of which maximally contains $Q$.
Based on earlier results~\cite{Alexe2011,Foniok2008}, 
ten Cate and Dalmau show that the frontier of $Q$ exists 
(i.e., is finite) if and only if the core of $Q$ is $c$-acyclic
(a condition based on the incidence graph of $Q$ stating that
each cycle must contain at least one distinguished variable).
However, the queries making up the frontier of a $c$-acyclic query
may not themselves be $c$-acyclic, which means that we cannot use 
$c$-acyclicity of queries to build a finite structure based on maximal containment.

In more recent work, ten Cate and colleagues have studied containment-based CQ fitting, generalization, specialization, and repair problems based on labeled examples~\cite{Cate23,Cate25}.

Barcel\'o et al.\ consider the problem of, given a CQ $Q$, find an approximation $Q'$ of $Q$
(i.e, $Q'$ is maximally contained in $Q$)
such that $Q'$ is in a class of CQs whose evaluation can be performed efficiently
in terms of combined complexity (e.g., acyclic CQs)~\cite{Barcelo14}.
In a sense, each query $Q'$ in our set of maximal restrictions of a query $Q$ 
(which comprises all queries maximally contained in $Q$) is an approximation of $Q$.  
However, for us, $Q$ and $Q'$ are always in the same class of CQs, namely 2CQs.
Some bound on the number of repeated occurrences of a relation symbol in a CQ
is required, because otherwise there may be no (finite) CQ maximally contained in $Q$,
even if $Q$ is a 2CQ, as we show in Section~\ref{sec:non-existence}.

Finally, efficient solutions for proving query (in)equivalence have been the focus of recent investigations, e.g., 
\cite{chu2018,haynes,He2024}.


\section{Discussion and next steps}
\label{sec:conclusions}

We have developed the first semantic metric for relational database queries and established the complexity of computing this metric.  This study was motivated by the diversity of practical applications of semantic query distance in data management.
Two natural next steps are (1) the algorithmics and empirical study of computing $\distCont$; and (2) investigating  
what happens in the presence of integrity constraints \cite{BarceloFGP20,BarceloPR17}.


\bibliographystyle{plain}
\bibliography{references}

\end{document}